\documentclass[twocolumn,showpacs,amsmath,amssymb,prl,superscriptaddress,floatfix]{revtex4}

\usepackage{graphicx,amssymb,amsmath}

\def \KK {\mathcal{K}}

\def \CCC {\mathbb{C}}
\def \EEE {\mathbb{E}}

\def \equals {\ = \ }

\begin{document}
\author{Alex D. Gottlieb}
\email{www.alexgottlieb.com} \affiliation{Wolfgang Pauli
Institute, Nordbergstrasse 15, A-1090 Vienna, Austria}
\author{Thorsten Schumm}
\affiliation{Atominstitut der \"Osterreichischen Universit\"aten,
TU-Wien, Stadionallee 2, A-1020 Vienna, Austria}
\title{Opposite sign correlations in fermion or boson gases}
\begin{abstract}

We investigate pair correlations in trapped fermion and
boson gases as a means to probe the quantum states producing the density fluctuations.   
We point out that ``opposite sign correlations" (meaning pair correlations that are positive for fermions and negative for bosons) 
unambiguously indicate that the quantum many-particle state cannot be ``free."
In particular, a system of fermions that exhibits positive pair correlations
cannot be described by any Slater determinant wavefunction. 
This insight may help one to interpret results of 
current experiments on ultracold atomic gases.

\end{abstract}

\maketitle

In 1956 Hanbury Brown and Twiss performed a landmark experiment
where they observed photon bunching in light emitted by a chaotic
source \cite{HBT_experimental}. The correlations they detected
% were a sign of constructive two-particle interference by bosons \cite{Fano} and 
``stimulated the birth of modern quantum optics" 
\cite{boson_bunching, fermion_bunching}. 
Forty years later, in another pioneering experiment,
temporal Hanbury Brown-Twiss (HBT) correlations were observed for
massive bosonic particles in a cold atomic beam, followed by the
observation of antibunching in electron and neutron beams
\cite{YasudaShimizu,KiesselRenzHasselbach,Neutron_antibunching}.
Bunching (or antibunching) may now be observed directly in
position space when a boson (or fermion) gas is released from a
trap \cite{boson_bunching,fermion_bunching}.

The last twelve years have witnessed rapid advances in the field
of ultracold atoms, and there has naturally been great interest in
the density correlations that can be measured 
in experiments on quantum gases of these atoms
\cite{% dalibard_castin,
NaraschewskiGlauber,GrondalskiAlsingDeutsch}.
Nowadays (2004-2007) correlations are being used to investigate
ever more complex quantum states, e.g., squeezed number states of
systems of indistinguishable atoms going through the Mott
insulator transition in optical lattices \cite{Foelling et al,Rom
et al}. Momentum correlations are studied in collisions of two
Bose-Einstein condensates and in the dissociation of weakly bound
dimers \cite{Perrin, Greiner et al}. In weakly interacting low
dimensional systems, where no clear phase transition to a
condensed state takes place at low temperatures, correlations are
used to characterize the degeneracy of the system \cite{Esteve et
al}. In the strongly interacting regime, second- and third-order  
correlations at zero distance have been used to identify the
Tonks-Girardeau gas \cite{Weiss_tonks-correlations,Phillips_tonks}.

One may analyze pair correlations in order to learn about the quantum states of trapped boson gases or atoms in optical lattices \cite{NaraschewskiGlauber,GrondalskiAlsingDeutsch}.   The power of this kind of intensity interferometry as a tool to probe 
complex many-body quantum states of ultracold atoms 
is highlighted in a short 
article published in 2004, which considers models of superfluid
fermion gases and optical lattice Mott insulators 
\cite{AltmanDemlerLukin}.  These models feature what we are calling opposite sign correlations, i.e.,  
negative pair correlations for bosons, whose
symmetry index is conventionally positive, and positive pair correlations for fermions, whose symmetry index
is conventionally negative.

Our purpose here is to point out that
if a many-boson or many-fermion system exhibits opposite sign correlations, then the state in question necessarily has a certain complexity. For
example, consider a fermion gas of --- say --- $10^5$  atoms.  If 
the gas exhibits any positive pair correlations when it has been prepared in a certain state, then that 
state cannot be represented by a simple Slater determinant wave function. In
general, if one probes a many-boson or many-fermion state and
finds that it exhibits opposite sign correlations, then, even
without any model for the unknown state, one may infer that it
is not a ``free" state, i.e., it does not have the form of a
grand canonical ensemble for noninteracting indistinguishable
particles.   We believe that opposite sign correlations can be
observed in current experimental setups and may even have already
been observed and passed unnoticed \cite{footnote}.

Let us discriminate between correlations that are merely due to
Bose or Fermi statistics, and correlations that reflect a further
complexity of the quantum state, the kind of complexity that
demands for itself a {\it post}-Hartree or {\it post}-Hartree-Fock
description. The bunching or antibunching observed in HBT effects
\cite{YasudaShimizu,KiesselRenzHasselbach,boson_bunching,fermion_bunching,Foelling
et al, Rom et al} can be attributed to the symmetric or
antisymmetric character of the many-particle wave functions that
describe the system.  For example, fermion antibunching can be
explained if one considers the simplest many-fermion wavefunctions, 
i.e., Slater determinants, which are just
antisymmetrized products of one-particle wave functions. Slater
determinants necessarily feature the negative correlations that
one sees in the (fermionic) HBT effect. However, more complex
many-fermion wave functions, which must be represented by {\it
superpositions} of two or more Slater determinants, can exhibit
{\it positive} correlations.  For example, let
$|1\rangle,|2\rangle,|3\rangle,$ and $|4\rangle$ represent four
orthonormal one-fermion states and let $|1\rangle \wedge
|2\rangle$ represent the two-particle Slater determinant formed
from states $|1\rangle$ and $|2\rangle$.  The two-particle state
\[
    \psi \equals \tfrac{1}{\sqrt{2}}\  |1\rangle \wedge |2\rangle  \ +\  \tfrac{1}{\sqrt{2}}\ |3\rangle\wedge |4\rangle 
\]
does involve positive correlations of occupation number: occupation of state $|1\rangle$ is perfectly correlated with occupation of state $|2\rangle$.  
This state features correlations
that cannot be explained by mere antisymmetry because $\psi$
cannot be written as a single Slater determinant.  To distinguish
these kinds of correlations from simple (HBT) correlations, we
refer to  Slater determinant wave functions as ``free" states and
we call complex states like $\psi$ ``nonfree" \cite{GottliebMauser1,GottliebMauser2}.   
When quantum chemists and condensed matter physicists speak of ``correlated"
systems of electrons, they usually mean ``nonfree" systems.

A free state is a state of a fermion or boson field in which
occupation numbers, relative to some system of orthogonal one-particle modes, 
are statistically independent.    
Slater determinant wavefunctions represent free states since the one-particle orbitals that participate in the determinant are certainly occupied,  
while any orbital orthogonal to these is certainly unoccupied.   The other fermionic free states are best represented by density matrices on the Fock space.  
All grand canonical thermal equilibrium ensembles of noninteracting fermions at positive temperature are examples of the latter.      All number states in the fermion Fock space are free states, but the only {\it pure} bosonic free state is the vacuum; all others are {\it mixed} states, which have to be represented by density matrices of rank greater than $1$.  Bosonic free states are called ``gaussian" in the context of photon statistics,  
and are used to describe chaotic thermal radiation fields in quantum optics \cite{KlauderSudarshan}.

  A free state is completely determined by its one-particle correlation operator
(i.e., all first-order correlations, including off-diagonal
correlations).  
Free states are conserved by free dynamics (where the particles do not interact with one another)
and by efficient measurement of a one-particle observable.
Therefore, if an ultracold gas is prepared in a free state and subjected to 
a time-dependent external potential, then, as long as particle-particle interactions may be disregarded, 
the state of the gas will remain free.   
In particular, switching off a trapping potential and allowing a gas to undergo
``time-of-flight" should not destroy the freeness of its state.

The next few paragraphs only discuss systems of fermions; we'll
come back to bosons later.

If a many-fermion system is in a free state, and one then observes
the positions of all the particles, the random pattern or
``configuration" of points that results is a sample of a
``determinantal point process" \cite{GottliebDeterminantal}.  
Determinantal point processes crop up unexpectedly in diverse
areas of mathematics and physics, e.g., random matrix theory
\cite{Soshnikov,Johansson} and graph theory \cite{Lyons}, and they
have been studied extensively
\cite{Macchi,DaleyVere-Jones,Soshnikov,Lyons,ShiraiTakahashi,HoughKrishnapurPeresVirag}.
Although ``it remains to investigate the processes thoroughly in
connection to statistical inference" \cite{MollerWaagepetersen},
determinantal point processes have statistical properties
that can provide tests of ``determinantality" and, hence, of
``freeness." 
Specifically, determinantal point processes have the property that  
density fluctuations in disjoint regions are never positively correlated, and therefore, when    
an experiment on ultracold fermions reveals positive density-density correlations, 
this can only mean that the state of the many-fermion system is not free.

A {\it point process} on a bounded region of space $S$ is a random finite subset $X$ of $S$.
A point process $X$ is {\it determinantal
on} $S$ {\it with kernel} $\KK:S \times S
\longrightarrow \CCC$ (which is assumed to be self-adjoint) if 
\begin{equation}
\label{moments}
    \EEE\Big[ \prod_{j=1}^m \#(X \cap E_j) \Big]
\end{equation}
--- the statistical expectation of the product of the occupation counts for arbitrary disjoint measurable subsets $E_1,\ldots,E_m$ of $S$ --- 
equals
\begin{equation}
\label{kernel}
    \int_{E_1}\cdots\int_{E_m}  \det \big( \KK(x_i,x_j ) \big)_{i,j=1}^{\ m} dx_1 \cdots dx_m
\end{equation}
for all $m\ge 1$
\cite{DaleyVere-Jones,HoughKrishnapurPeresVirag}.  From
$(\ref{moments}) = (\ref{kernel})$ it follows that,
for a determinantal point process $X$, the numbers of
points in any two disjoint regions of space are never positively
correlated.  That is, if $R_1$ and $R_2$ are two disjoint regions
of space, and if $N_1=\#(X \cap R_1)$ and $N_2=\#(X \cap R_2)$
denote, respectively, the random number of points in each of these
regions, then $\langle N_1 N_2 \rangle \le \langle
N_1\rangle\langle N_2 \rangle$.  Furthermore, in case the particle
numbers in $R_1$ and $R_2$ are {\it uncorrelated}, i.e., when
$\langle N_1 N_2 \rangle = \langle N_1\rangle\langle N_2 \rangle$,
the two restricted point processes $X \cap R_1$ and $X \cap R_2$
are {\it independent}.   If one views a determinantal process
through an ``observation window" $R$ \cite{MollerWaagepetersen},
the restricted point process $X \cap R$ is also determinantal.

The configuration statistics of a free fermion state $\omega$ are determinantal with kernel
\[
     \KK(x,y) \equals \omega\big({\hat{\psi}}^{\dagger}(y)\hat{\psi}(x)\big),
\]
where $\hat{\psi}(x)$ denotes the field operator at $x$.  
The kernel of the point process viewed through observation window $R$ is simply
\[
    {\bf 1}_R(x) \KK(x,y) {\bf 1}_R(y) .
\]
Therefore, when one observes the configuration $X$ of a system of fermions in a free state $\omega$, the random occupation numbers
\begin{eqnarray*}
N_1 & = &  \#(X \cap R_1) \\
 N_2 & = & \#(X \cap R_2)
\end{eqnarray*}
satisfy
\begin{equation}
\label{antibunching}
\big\langle N_1 N_2  \big\rangle_{\omega} \le  \big\langle  N_1 \big\rangle_{\omega} \big\langle N_2 \big\rangle_{\omega}
\end{equation}
if $R_1$ and $R_2$ do not overlap.  More generally,
\[
\Big\langle \prod_{i=1}^k N_i \Big\rangle_{\omega} \le  \prod_{i=1}^k  \big\langle N_i \big\rangle_{\omega}
\]
if the the regions $R_1,R_2,\ldots,R_k$ are disjoint.  Violations of the ``antibunching" inequality (\ref{antibunching}) indicate that the many-fermion state $\omega$ is not free.  

For bosons, it is {\it negative} pair correlations that indicate nonfreeness.  A system of bosons cannot be in a free state if the density fluctuations at any pair of (distinct) points are negatively correlated.   The point processes corresponding to the bosonic free states are known as ``permanantal point processes"  \cite{Macchi,McCullaghMoller} and these feature positive density-density correlations.  That is, if $X$ is the random configuration of bosons in a free state $\omega$, and if  $N_1= \#(X \cap R_1)$ and $N_2= \#(X \cap R_2)$, then
\begin{equation}
\label{bunching}
\big\langle N_1 N_2  \big\rangle_{\omega} \ge  \big\langle  N_1 \big\rangle_{\omega} \big\langle N_2 \big\rangle_{\omega}
\end{equation}
whenever $R_1$ and $R_2$ disjoint, and therefore, violations of the ``bunching" inequality (\ref{bunching}) signify that the many-boson state $\omega$ is not free.  On a quantum statistical level, the HBT correlations that may be observed in starlight are due to the thermal character of the source of that light, from which gaussian (i.e., free) photon fields are assumed to emanate \cite{Glauber,KlauderSudarshan}.   Incoherent thermal sources produce radiation fields whose pair correlations are positive or zero, but other kinds of light sources can produce photon fields with {\it negative} pair correlations \cite{footnoteKlauderSudarshan}.

 Free states correspond to {\it grand canonical} ensembles of noninteracting bosons or fermions, but it is more appropriate to model thermal equilibrium states of ultracold gases by the {\it canonical} ensemble, for the trapped gas is a closed system of massive particles.   However, when the number of particles is large, it is reasonable to suppose that canonical ensembles of free bosons (or fermions) exhibit {\it nearly} permanantal (or determinantal) statistics.    Such ``equivalence of ensembles" has recently been established rigorously for untrapped particles \cite{TamuraIto}.

In conclusion, when opposite sign correlations are observed, this suggests that interactions between the particles, whether they be collisions or indirectly mediated interactions, are responsible for a certain structural complexity of the many-body state.   Such opposite sign correlations should be detectable in absorption images of ultracold atomic gases.  When opposite sign correlations are observed one may infer that the state in question is definitely not free.  Remarkably, this inference does not require any knowledge or model of the unknown state.

\bigskip

\noindent {\bf Acknowledgements.}
{\it  We thank A. Scrinzi and J. Schmiedmayer for very helpful instruction.  A.G. is supported by the
Austrian Ministry of Science via its grant for the
Wolfgang Pauli Institute and by the Austrian Science Foundation
(FWF) via the START Project Y-137-TEC.}


\begin{thebibliography}{10}
\bibitem{footnote}
We have in mind Refs. 9 and 10 (see Figure 4d in both Letters).
\bibitem{footnoteKlauderSudarshan}
See page 246 of Ref. 20, where this is explicit.
\end{thebibliography}
\end{document}